\documentclass[10pt,fleqn]{article}

\usepackage[breaklinks]{hyperref}
\usepackage[german,british]{babel}

\usepackage{fancyheadings}

\usepackage{amsmath}
\usepackage{amssymb}

\usepackage{graphicx}
\usepackage{psfig}
\usepackage{color}
\usepackage{epsfig}

\newcommand{\dif}{\mathrm{d}}

\DeclareFontFamily{OT1}{rsfs}{}
\DeclareFontShape{OT1}{rsfs}{m}{n}{ <-7> rsfs5 <7-10> rsfs7 <10->rsfs10}{} 
\DeclareMathAlphabet{\mycal}{OT1}{rsfs}{m}{n}

\begin{document}
\title{The Einstein static universe with torsion 
and the sign problem of the cosmological constant}
\author{C G B\"ohmer\footnote{E-mail: boehmer@hep.itp.tuwien.ac.at}}
\date{}
\maketitle

%%% TUW preprint number on titlepage
\thispagestyle{fancy}
\setlength{\headrulewidth}{0pt}
\rhead{TUW--03--31, Preprint ESI 1384}
\noindent
\textit{Institut f\"ur Theoretische Physik,
Technische Universit\"at Wien, Wiedner Hauptstrasse 8-10, A-1040 Wien, \"Osterreich.}
\\ \mbox{} \\
\begin{abstract}
In the field equations of Einstein-Cartan theory with cosmological 
constant a static spherically symmetric perfect fluid with spin density 
satisfying the Weyssenhoff restriction is considered. This serves as
a rough model of space filled with (fermionic) dark matter.
From this the Einstein static universe with constant torsion is
constructed, generalising the Einstein Cosmos to Einstein-Cartan theory.

The interplay between torsion and the cosmological constant is discussed. 
A possible way out of the cosmological constant's sign problem is
suggested.
\end{abstract}
\mbox{} \\ 
\mbox{} 

\newpage
\section{Introduction}
Cosmological observations~\cite{Krauss:2003yb,Perlmutter:1998zf} 
give strong indications of 
the presence of a positive cosmological constant, which would mean 
that the universe is of
\mbox{de Sitter} type. On the other hand the low energy limit of 
supersymmetry theories prefers a negative cosmological constant, 
implying an anti-de Sitter 
cosmos~\cite{VanNieuwenhuizen:1981ae}. Existing solutions to this problem necessitate
the inclusion of a further field (quintessence)~\cite{Carroll:1998zi}. 

It is shown that this may not be necessary.
By considering the Einstein-Cartan theory with cosmological constant a 
model is constructed which indicates that a relatively simple solution 
to the problem may be possible. 

\section{Field equations in Einstein-Cartan theory}
In the Cartan formalism the metric is expressed in terms of
vielbein 1-forms $e^i$
\begin{align}
      \dif s^2 = \eta_{ij}\, e^i \otimes e^j,
      \label{metric}
\end{align}
indicating anholonomic indices.
In standard notation torsion and curvature 2-forms are given by
\begin{align}
      &T^i = (De)^i =\dif e^i + \omega^i{}_j \wedge e^j
      =\frac{1}{2}T^i{}_{jk}\, e^{j}\wedge e^{k}, 
      \label{torsion}\\
      &R^i{}_j= (D\omega)^i{}_j=\dif\omega^i{}_j+
      \omega^i{}_k \wedge\omega^k{}_j
      =\frac{1}{2}R^i{}_{jkl}\, e^k \wedge e^l,
      \label{curvature}
\end{align}
respectively, where $\omega^i{}_j$ is the connection 1-form. 
The field equations of the Einstein-Cartan theory are obtained
from the action 
functional~\cite{Gonner:1984rw,Hehl:1973,Hehl:1974,Hehl:1976kj,Prasanna:1975wx}
\begin{align}
      S = \int \bigl(L+2\Lambda\mathbf{\epsilon}+\kappa L_m \bigr),
      \label{action}
\end{align}
where $L=\frac{1}{2}R \mathbf{\epsilon}$ and 
$L_m =L_m(e^i,\omega^i{}_j)$ is the matter Lagrangian.
$\mathbf{\epsilon}$ is the volume four form, $R=\eta^{ln}\delta^m_k R^k{}_{lmn}$
is the Ricci scalar and $\kappa$ the gravitational coupling constant.

Variation of~(\ref{action}) with respect to $e^i$ and
$\omega^i{}_j$ together with~(\ref{torsion}) and~(\ref{curvature}) 
yields the field equations in Einstein-Cartan 
theory~\cite{Hehl:1974cn,Hehl:1976kj,Prasanna:1975wx}
\begin{align}
      &R^i{}_j - \frac{1}{2}R \delta^i_j + \Lambda\delta^i_j = -\kappa t^i{}_j 
      \label{ec_ric}\\
      &T^i{}_{jk}-\delta^i_j T^l{}_{lk}-\delta^i_k T^l{}_{jl}= -\kappa s^i{}_{jk},
      \label{ec_tor}
\end{align}
where $t^i{}_j$ is the canonical energy-momentum tensor and 
$s^i{}_{jk}$ is the tensor of spin.

\section{Static spherically symmetric spacetime}
A static and spherically symmetric spacetime 
in the field equations~(\ref{ec_ric}) and~(\ref{ec_tor})
is described by a line element of the form
\begin{align}
      \dif s^2 = A^2 \dif t^2 - B^2 \dif r^2 - r^2 \dif \Omega^2,
      \label{sssmetric}
\end{align}
with $A,B$ functions of the radial coordinate $r$ only.
If one assumes that the spacetime is filled with a fermionic fluid,
a classical description of the spin contribution in~(\ref{ec_tor}) is
\begin{align}
      s^i{}_{jk} = u^i S_{jk},\qquad u^i S_{ji} = 0,
      \label{spin}
\end{align}
where $u^i$ is the four velocity of the fluid and $S_{ij}$ is the intrinsic
angular momentum tensor. Further restricting to an isotropic perfect 
fluid~\cite{Kopczynski:1973} yields
\begin{align}
      t_{ij} &= h_i u_j - P \eta_{ij}, \nonumber \\
      h_i &= (\rho +P)u_i - u^j \nabla_k(u^k S_{ji}) \nonumber \\
          &= (\rho +P)u_i + a^j S_{ji},
      \label{em}
\end{align}
where $\rho$ is the matter density and $P$ the isotropic pressure,
$a^j=(u^k \nabla_k)u^j$ is the fluid's acceleration. Thus the
canonical energy-momentum tensor is symmetric if the acceleration of
the fluid is zero~\cite{Griffiths:1982}. 

Assuming spherical symmetry for the spin implies that $S_{ij}$ has only
one non-vanishing component~\cite{Kopczynski:1972}, $S_{23}=K$, where
$K$ is a function of $r$. Since a static 
configuration is also assumed $u^i=\delta^i_0$ and hence $s^0{}_{23} = K$.
Therefore, one can solve~(\ref{ec_tor}) and one finds
\begin{align}
      T^0{}_{23}=-T^0{}_{32}=-\kappa K, 
      \label{tor}
\end{align}
all other components vanish.
Equation~(\ref{torsion}) implies that $T^0=-\kappa K e^2 \wedge e^3$,
and $T^1=T^2=T^3=0$. Taking~(\ref{spin}) into account~(\ref{em})
simplifies to $t^0{}_0=\rho$ and $t^1{}_1=t^2{}_2=t^3{}_3=-P$.  

The remaining field equations~(\ref{ec_ric}) are three independent 
equations which imply energy-momentum (plus spin) conservation. 
For convenience we use the first two field equations and
the conservation equation. With $\kappa=-8\pi$ this 
yields~\cite{Prasanna:1975wx,Som:1981sd}
\begin{align}
      &\frac{1}{r^2}\frac{\dif}{\dif r}\left(r-\frac{r}{B^2}\right)
      -\Lambda + 16\pi^2 K^2 = 8\pi \rho,
      \label{f1}\\
      &-\frac{1}{r^2}+\frac{1}{B^2}\left(\frac{2A'}{Ar}+\frac{1}{r^2}\right)
      +\Lambda + 16\pi^2 K^2 = 8\pi P,
      \label{f2}\\
      &P'+(\rho+P)\frac{A'}{A}-4\pi K\left(K'+K\frac{A'}{A}\right) = 0,
      \label{con}
\end{align}
where the prime denotes differentiation with respect to $r$.
Furthermore if we {\em assume} that the conservation equation 
\begin{align}
      P'+(\rho+P)\frac{A'}{A}=0,
\end{align}
of general relativity holds then~(\ref{con}) implies
\begin{align}
      K'+K\frac{A'}{A} = 0,\qquad K \propto A^{-1},\qquad\mbox{for\ } K \neq 0.
\end{align}
One may redefine~\cite{Hehl:1974,Prasanna:1975wx} the pressure and the energy 
density by
\begin{align} 
      &\rho_{\mbox{eff}}=\rho -2\pi K^2 +\frac{\Lambda}{8\pi},
      \label{rhoeff} \\
      &P_{\mbox{eff}}=P -2\pi K^2 -\frac{\Lambda}{8\pi},
      \label{peff}
\end{align}
and rewrite equations~(\ref{f1})-(\ref{con}).
This leads to the usual field and conservation equations with
vanishing torsion and vanishing cosmological constant. 

The effect of the cosmological constant can be seen 
as a special type of an energy-momentum tensor. It acts as an
unusual fluid with $P^{\Lambda}= -\Lambda/8\pi$ and 
$P^{\Lambda}=-\rho^{\Lambda}$ as an equation of state. 
On the other hand, the torsion contribution $K$ acts as a fluid 
with $P^{K}= -2\pi K^2$ and equation of state $P^{K}=\rho^{K}$.
For simplicity from now on $\rho$, $P$ and $K$ are assumed to be constant.

It is instructive to have a closer look at the  
effective quantities~(\ref{rhoeff}) and~(\ref{peff}). The
cosmological solution consisting of an incoherent dust $P=0$
with vanishing torsion $K=0$ and \mbox{de Sitter} type universe implies
$\rho_{\mbox{eff}}>0$ and $P_{\mbox{eff}}<0$. 

Thus one can try to reconcile this with a negative cosmological constant
as required from supergravity~\cite{VanNieuwenhuizen:1981ae}.
Two assumptions are needed. (i) Cosmological observations measure the 
effective energy density and pressure and from this the sign of the 
cosmological constant is defined. Note that this is a crucial
assumption for what is argued in the following. 
\mbox{(ii) The} output of supersymmetry 
theories is correct in the sense that the cosmological constant is negative.

Then one can check whether the three conditions $|\Lambda|=-\Lambda>0$,
$\rho_{\mbox{eff}}>0$ and $P_{\mbox{eff}}<0$ can be satisfied
simultaneously.
\begin{align}
      \rho_{\mbox{eff}}&>0 \Rightarrow \rho-\frac{|\Lambda|}{8\pi}>2\pi K^2, \\
      P_{\mbox{eff}}&<0 \Rightarrow P+\frac{|\Lambda|}{8\pi}<2\pi K^2.
\end{align}
Assuming as before an incoherent dust $P=0$, the above leads to the inequality
\begin{align}
      \frac{|\Lambda|}{8\pi}<2\pi K^2<\rho-\frac{|\Lambda|}{8\pi}.
      \label{ineqn}
\end{align}
If the (AdS) cosmological constant has an upper bound given by torsion and if
the cosmic energy density is sufficiently large then the three conditions
can simultaneously be satisfied. 
From this one can conclude that under the above assumptions it is 
possible to reconcile observational data leading to a positive cosmological
constant, and the supersymmetry requirement yielding a negative sign. 
The argument also works for vanishing cosmological constant.

For easier comparison with observational data the inequality~(\ref{ineqn}) is
divided by the critical density $\rho_c$ and rewritten in terms of density
parameters $\Omega$. This gives
\begin{align}
      \Omega_{|\Lambda|}^{\mbox{susy}} < \frac{1}{\rho_c}
      \Bigl(\frac{K}{2M_{\mbox{pl}}}\Bigr)^2 <
      \Omega - \Omega_{|\Lambda|}^{\mbox{susy}},
\end{align}
where the notation of~\cite{Liddle:2000cg} was used. The present
value of the dark matter contribution is denoted 
by $\Omega_0$, recent observations 
suggest~\cite{Krauss:2003yb} that $\Omega_0=0.3\pm0.1$. Thus for small enough
$\Omega_{|\Lambda|}^{\mbox{susy}}$ one gets an upper and a lower 
bound for the torsion contribution.

If the second assumption is dropped (no supersymmetry) and if one 
considers a positive cosmological constant it is found that observations are
compatible with the presence of torsion. In this case from~(\ref{rhoeff}) 
and~(\ref{peff}) one can deduce
\begin{align}
      \Omega + \Omega_{\Lambda} >\frac{1}{\rho_c} 
      \Bigl(\frac{K}{2M_{\mbox{pl}}}\Bigr)^2 \ge 0,
\end{align}
that the density parameter of torsion is heavily constrained.
Observations are, of course, also compatible with vanishing torsion.

\section{Einstein universe with constant torsion}
Finally the Einstein universe with constant torsion is constructed.
The first field equation~(\ref{f1}) can easily be integrated and 
yields
\begin{align}
      \frac{1}{B^2} = 1 - \frac{2 \mathcal{M}(r)}{r}
      -\frac{\Lambda}{3}r^2 + \frac{\mathcal{K}(r)}{r},
      \label{b}
\end{align}
where the constant of integration was set to zero because
of regularity at the centre and
\begin{align}
      \mathcal{M}(r) = \int_0^r 4\pi \rho s^2 \dif s,\qquad
      \mathcal{K}(r) = \int_0^r 16\pi^2 K^2 s^2 \dif s.
\end{align}
From equations~(\ref{f2}) and~(\ref{con}) one 
can eliminate $A'/A$ and gets the
Tolman-Oppen\-heimer-Volkoff 
equation~\cite{Oppenheimer:1939ne,Tolman:1939} with cosmological
constant and spin contribution
\begin{align}
      P' = -r\frac{\bigl(
      12\pi P+4\pi \rho_0 -\Lambda -32\pi^{2}K_0^2\bigr)
      \bigl(P+\rho_0 -4\pi K_0^2 \bigr)}
      {3-\bigl(8\pi \rho_0 +\Lambda -16\pi^2 K_0^2 \bigr)r^2},
      \label{pressure}
\end{align}
where we assumed positive constant density 
$\rho = \rho_0 = \mbox{const.}$ and positive
constant torsion $K=K_0=\mbox{const.}$ 

If $P'=0$ for all $r$ the differential 
equation~(\ref{pressure}) of the pressure implies
\begin{align}
      \Lambda = 4\pi \bigl(3P_0 + \rho_0 -8\pi K_0^2 \bigr),
      \label{lambda_e}
\end{align}
provided the modified energy condition
\begin{align}
      P_0 + \rho_0 -4\pi K_0^2 > 0,
\end{align}
holds. Moreover, under these assumptions~(\ref{con}) implies
that $A=\mbox{const.}$ which we can re-scale to one.

The metric function $B$ can now be read from~(\ref{b}). We also
insert the value of the cosmological constant~(\ref{lambda_e}) 
and arrive at
\begin{align}
      \frac{1}{B^2}= 1 - 4\pi\left(\rho_0 + P_0 -4\pi K_0^2\right)r^2.
\end{align} 
From this one can read off the radius of the Einstein static universe
with torsion
\begin{align}
      R_E^2 = \frac{1}{4\pi(\rho_0 + P_0 -4\pi K_0^2)},
\end{align}
which for vanishing $K_0$ coincides with the radius of the standard Einstein
static universe.

The above solution is a modified Einstein static universe. The
modification is due to the additional spin contribution $K$.
Of course for $K=0$ the cosmological constant of the usual
Einstein static universe~\cite{Einstein:1917ce} is 
reproduced by~(\ref{lambda_e}). Torsion free generalisations of the 
Einstein static universe have been published earlier in 
Ref.~\cite{Ibrahim:1976}\footnote{I would like to thank Aysel 
Karafistan for bringing this reference to my attention.} and 
also in Ref.~\cite{Boehmer:2002gg}.

\section{Conclusions and Outlook}
Under the assumptions (i) and (ii) it was possible to
construct a model in which the different observational (dS) and supersymmetry (AdS)
requirement concerning the cosmological constant's sign could be incorporated. 

It would be interesting to find a model in which one could clearly
define an effective cosmological constant and therefore solving the 
full problem. Other sources of torsion might be a good starting point.

A detailed investigation of observational data along the lines 
of the recent report~\cite{Hammond:2002rm} is beyond the scope of the 
present work.

The static Einstein Cosmos was generalised to Einstein-Cartan theory.
A possible generalisation of the present approach is to weaken to
strong Weyssenhoff condition and consider instead a 
hyperfluid~\cite{Hehl:privat,Obukhov:1993pt}. The qualitative result that
the Einstein Cosmos can be generalised will be unaffected by that. 

\subsection*{Acknowledgements}
I would like to thank Daniel Grumiller and Herbert Balasin for constant help and 
advice and Wolfgang Kummer for suggestions.

I am grateful to Friedrich W. Hehl for the stimulating discussions and useful 
comments during the ESI Workshop on Gravity in Two Dimensions.

\addcontentsline{toc}{section}{References}
\bibliographystyle{plain}
\bibliography{review}

\end{document}